\documentclass{aa}
\usepackage{graphicx}
\usepackage{txfonts}
\usepackage{natbib}
\def\approxgt{\mathrel{\hbox{\rlap{\lower.55ex \hbox {$\sim$}}
        \kern-.3em \raise.4ex \hbox{$>$}}}}
\def\approxlt{\mathrel{\hbox{\rlap{\lower.55ex \hbox {$\sim$}}
        \kern-.3em \raise.4ex \hbox{$<$}}}}

\def\chandra {\emph{Chandra }}
\def\chandran {\emph{Chandra}}
\def\xmmn {\emph{XMM-Newton }}

\def\pn  {\emph{pn }}
\def\pnn {\emph{pn}}
\def\mos  {\emph{MOS }}
\def\mosn {\emph{MOS}}
\def\mn  {\emph{MOS1 }}
\def\md  {\emph{MOS2 }}
\def\mun {\emph{MOS1}}
\def\mdn {\emph{MOS2}}
\def\acis  {\emph{ACIS }}
\def\acisn {\emph{ACIS}}
\def\st {\emph{ACIS-S3 }}
\def\stn {\emph{ACIS-S3}}
\def\epic  {\emph{EPIC }}
\def\epicn {\emph{EPIC}}

\newcommand{\ciao }{{\em CIAO}}
\newcommand{\heasoft }{{\em Heasoft}}

\begin{document}
   \title{Non-thermal emission in the core of Perseus: \\
results from a long \xmmn observation}
   \titlerunning{A study of non-thermal emission in Perseus with \xmmn}
 \author{Silvano Molendi
   \inst{1}
   \and
    Fabio Gastaldello
   \inst{2,3,4}
          }

\institute{
          IASF-Milano, INAF, via Bassini 15,
         I-20133 Milano, Italy
        \email{silvano@iasf-milano.inaf.it}
        \and
         Dip. di Astronomia, Universit\`a di Bologna,
         via  Ranzani 1,
         I-40127 Bologna, Italy
         \email{fabio.gastaldello@unibo.it}
         \and
         INAF, Osservatorio Astronomico di Bologna, via
         Ranzani 1, Bologna 40127, Italy
         \and
         Department of Physics and Astronomy, University of California at Irvine, 4129
         Frederick Reines Hall, Irvine, CA 92697-4575
         }

   \date{To appear in Astronomy \& Astrophysics Main Journal}

\abstract{ We employ a long \xmmn observation of the core of the
Perseus cluster to validate claims of a non-thermal component
discovered with \chandran. From a meticulous analysis of our
dataset, which includes a detailed treatment of systematic errors,
we find the 2-10 keV surface brightness of the non-thermal component
to be smaller than about
5$\times10^{-16}$erg$~$cm$^{-2}$s$^{-1}$arcsec$^{-2}$. The most
likely explanation for the discrepancy between the \xmmn and
\chandra estimates is a problem in the  effective area calibration
of the latter. Our \epic based magnetic field lower limits are
not in disagreement with Faraday rotation measure estimates on a few cool
cores and with a minimum energy estimate on Perseus. In the not too
distant future \emph{Simbol-X} may allow detection of non-thermal
components with intensities more than 10 times smaller than those
that can be measured with \epicn; nonetheless even the exquisite
sensitivity within reach for \emph{Simbol-X} might be insufficient
to detect the IC emission from Perseus.

   }
\maketitle
\section{Introduction} \label{sec: intro}


Although the bulk of the energy radiated from clusters is of thermal
nature, non-thermal mechanisms play an important role. Indeed the
characterization of non-thermal components provides much needed
clues on the physical process presiding over the formation and
evolution of clusters.
In some of the more disturbed objects, evidence of non-thermal
processes has been known for quite some time. Radio observations
indicate that merging clusters are often the site of cluster-wide
synchrotron emission, the so called radio halos and radio relics
\citep[and references therein]{cassano07}. Radio data, polarimetric
or not, has been used to provide estimates of magnetic fields in
clusters. Unfortunately estimates based on Faraday rotation measures
(hereafter RM) or minimum-energy arguments are affected by large
uncertainties. In the case of RM estimates the unknown field
topology \citep{ensslin03} and the accessability of only a few, not
necessarily representative, lines of sight \citep{rudnick03}, are a
major source of concern, while for minimum energy arguments the
proton to electron ratio, and  the applicability of the argument
itself play an equally critical role.

It has been recognized for quite some time that detection of inverse
Compton (hereafter IC) emission at X-ray wavelengths can provide an
alternative method to estimate cluster magnetic fields
\citep{rephaeli87}. Detection and characterization of the so called
hard tails is by no means a trivial task: the signal, if there is
one, is caught between the hammer of the thermal emission and the
anvil of the instrumental background. So far, only detections at a
few sigma level have been  reported \citep[for a recent review
see][]{rephaeli08}, and at least in one case, Coma \citep{fusco99,fusco04,fusco07},
they have been challenged \citep{rossetti04,rossetti07}.

\begin{figure}
  \centering
  \resizebox{75mm}{!}{\includegraphics[angle=-90]{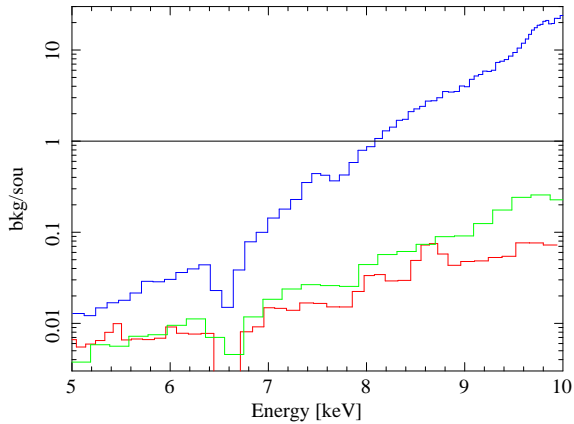}} \\
  \caption{Ratio of background to source spectra for the \st (blue),
  \md (green) and \pn (red) instruments from the
  1$^{\prime}$-2$^{\prime}$ ring. The \st ratio rises rapidly from 1\% at 5keV
  to unity at 8 keV; at 10 keV the \st background is about 20 times larger than
  the source. \mos and \pn ratios rise above 1\% around 7 keV and remain below
  30\% and 10\% respectively at all energies.}
  \label{fig: cxo_epc}
\end{figure}

Aim of the present work is to employ a long \xmmn observation of the
Perseus core to validate claims about a non-thermal emission
component discovered with \chandra \citep{sand04,sand05,sand07}. The
major advantage of the \xmmn observation is the better sensitivity
at high energies ($\sim$ 6 keV) of the \epic detectors with respect
to the \stn ; this is illustrated in Fig.~\ref{fig: cxo_epc} where
we show the ratio of background to source intensity in a
representative region, a ring with bounding radii
1$^{\prime}$-2$^{\prime}$ centered on the emission peak. The
\chandra ratio rises very rapidly from 1\% at 5keV to unity at 8
keV; at 10 keV the \st background is about 20 times larger than the
source. \mos and \pn ratios rise above 1\% around 7 keV and remain
below 30\% and 10\% respectively at all energies.

The outline of the paper is as follows: in Sect.~\ref{sec: ev_prep}
we provide details on our data preparation; in Sect.~\ref{sec:
spec_anal} we describe results from the spectra analysis; in
Sect.~\ref{sec: chandra} we compare results from the analysis of
\epic data with those from \chandran; in Sect.~\ref{sec: discussion}
we discuss our results and in Sect.~\ref{sec:
summary} we summarize them. We assume $H_0 = 70$~km
s$^{-1}$Mpc$^{-1}$ so that 1 arcsec$=$0.35 kpc at the redshift of
NGC1275 \citep[0.0176,][]{huchra99}; all errors are 1$\sigma$ unless
otherwise stated.

\section{Data preparation} \label{sec: ev_prep}

Perseus  was observed by \xmmn during revolution 1125 with the \epic
\mos (\pnn) detectors in Full Frame Mode (Extended Full Frame) with
medium filter on all \epic cameras. We have obtained calibrated
event files for the \mun, \md and \pn cameras using SAS version 7.1
`Emchain'', ``Epchain'' tasks and calibration files (CCFs) available
at the end of August 2007. A soft proton cleaning procedure was
performed by extracting a light curve in 100 second bins in the
10-12~keV energy band in a peripheral region of the \mos and \pn
cameras and applying a threshold of 0.20 cts s$^{-1}$ for \mos and
0.6 cts s$^{-1}$ for \pnn, to generate a filtered event list. The
total live time after screening is $1.12\times 10^5$ s for \mn and
\md and $6.6\times 10^4$ for \pnn. Finally, we filtered event files
according to \verb|FLAG|      (\verb|FLAG|==0) and \verb|PATTERN|
(\verb|PATTERN|$\leq$4 for \pn and \verb|PATTERN|==0 for \mosn). The
filtered event files, which were subsequently used for data
analysis, contained a total of about $16\times 10^6$, $19 \times
10^6$ and  $35 \times 10^6$ events for \mun, \md and \pn
respectively.

\subsection{Spectra preparation} \label{sec: spec_prep}

We have accumulated spectra from annuli with bounding radii
0.5$^\prime$-1$^\prime$, 1$^\prime$-2$^\prime$ and
2$^\prime$-3$^\prime$, and from sectors 45 degrees wide from the
same annuli.

As already pointed out in Sect.~\ref{sec: ev_prep}, for the \mos
detectors only \verb|PATTERN|==0 events were selected, this was done
to avoid pile up which, albeit small (few $\% $), can have
non-negligible consequences within $\sim$ 2$^\prime$ from the
cluster core \citep[i.e. be sufficiently strong to distort the spectral
shape, particularly at high energies,][]{mole03}. We generate flux weighted effective areas
using the \verb|arfgen| task with exposure corrected images of the
source and the parameter extended source switched to true. We also
correct \pn spectra for out of time events following the
prescriptions of \citet{grupe01}. Redistribution matrices were
generated with the \verb|rmfgen| task, one matrix was generated for
\mn and \md respectively, a set of 10 matrices were generated for
\pn to account for the dependence of the \pn spectral resolution
upon the distance to the camex.

Although Perseus is the brightest cluster in the X-ray sky and \epic
offers the largest effective area of any X-ray experiment flown thus
far, a reliable characterization of the background is one of the
fundamental requirements to carry out a search for non-thermal
emission. Our background subtraction procedure makes use of a sum of
blank fields selected by our own group \citep{leccardi08}. Spectra
have been accumulated from the same sky regions as the source
spectra, after re-projection onto the sky attitude of the source. To
allow for differences in the intensity of the background components
between source and background observations we re-normalize the
background spectra. The re-normalization constant is determined by
comparing the intensity of the total emission in source and
background observations from a peripheral region and in a hard
(10-12 keV for \mn and \mdn, 12-14 keV for \pnn) energy band where
source emission is expected to be small. The re-normalization
constants, $R_k$, are found to be 1.64, 1.57 and 1.52 for \mun, \md
and \pn respectively. To verify the correctness of our procedure we
performed a spectral fit to the data in the outer ring used to
determine the re-normalization constants, the fit was performed in a
hard, albeit source dominated band (4-7 keV). The best fitting model
was then extrapolated to a background dominated band (8-14 for \pn
and 8-12 for \mosn) where it was found to reproduce adequately the
source spectrum. Variations of the re-normalization constant
described above of only a few percent result in a best-fitting model
extrapolation significantly under or over-shooting the source
spectrum, indicating that our determination of the re-normalization
constants is good to a few percent. To account for possible
variations in the background intensity over the detectors and the
issuing re-normalization errors on the background subtraction  we
have assumed a 5\% systematic uncertainty on the re-normalization
constants we apply on all our source spectra. In practice this is
achieved by performing 3 different fits for each spectrum one with
the background with the standard re-normalization constant, $R_k$,
the other two with a larger,  $R_k + 0.05\times R_k$, and smaller,
$R_k - 0.05\times R_k$, re-normalization constants respectively.

As a further check of  the quality of our background subtraction
technique we have also pursued a different venue , i.e. background
modeling, following the lines described in a recent paper
\citep{leccardi08}. A comparison of results obtained with the two
methods shows that within some 4 arc-minutes of the cluster core
they provide virtually indistinguishable results. Given that the
background modeling alternative is more cumbersome from the point of
view of the spectral analysis and that it has only been developed
for the \mos detectors \citep{leccardi08} in the following we
report results obtained with the background subtraction method.










One of the hallmarks of the Perseus observations is its high
statical quality, typical spectra contain in the order of hundred of
thousands to millions of source counts. Under these extreme
conditions statistical errors can be minuscule, especially around 1
keV where effective areas peak: here systematic errors become an
important if not dominant component of the error budget.  This can
be readily verified by inspecting spectral fits,
typically residuals appear at the position of edges where rapid variations in
the effective areas occur. Conversely far less statistics is
available in the 5-10 keV band where some astrophysical features,
such as the thermal exponential cutoff, Fe and Ni emission lines are
located. The question is then how to prevent the $\chi^2$ fitting
procedure from giving more importance to statistically significant
spurious features located around 1 keV than to astrophysical
signatures at higher energies. We adopt two independent methods: the
first is to identify the cause of systematic errors  and evaluate
their impact on parameter estimates; the second is to associate a
systematic error to spectra. While the former strategy may be
considered as part of the modeling process of Perseus spectra, and
as such will be described in detail in the following section, the
latter is accomplished by associating a systematic error to all
spectral bins. We choose a somewhat moderate and energy independent value of 2\%  for the systematic error
for three reasons: firstly it is consistent with a conservative estimate for the relative calibration
of the \epic instruments \citep{kirsch07}; secondly it has the effect of increasing errors in
the low energy part of the spectrum to the point that they become comparable
to errors at higher energies;  thirdly a detailed energy dependent characterization of
systematic errors is in part achieved through the modeling processes
described in the following section and in part by associating the same
relative systematic error to spectral regions characterized by very different
statistical errors as already clarified in the previous point.
Inclusion of the systematic error is technically achieved
by modifying  the {\verb SYS_ERR } keyword in the source spectra header.
The XSPEC package associates to each spectral bin an error which is
the sum in quadrature of the statistical and  systematic error.

\section{Spectral analysis} \label{sec: spec_anal}
Due to its higher effective area we have considered  the \pn camera
as lead instrument in our analysis. Results from the analysis of
\mos data is also reported (see Sect.~\ref{sec: ring_1_2_mos}). We
have made use of two kind of baseline models: a two temperature plus
power-law model, in XSPEC jargon {\verb phabs*(2vmekal+pegpwrlw) }, which
we shall refer to as 2TP, and a multi temperature plus power-law
model, {\verb phabs*(5vmekal+pegpwrlw) }, hereafter MTP. For the former
model the free parameters are the N$_{\rm H}$, the temperatures, the
normalizations of the thermal and pow-law components, the redshift,
the O, Mg, Si,
S, Ar, Ca, Fe and Ni abundances. Ne, Na and Al abundances are tied
to O; metal abundances for a given species of one component are tied
to those of the other; the photon index of the power-law is fixed to
1.8, one of the values adopted in \citet{sand05} and \citet{sand07}.
For the latter model the free parameters are the N$_{\rm H}$,
the normalizations of the thermal and power-law components, the redshift,
the O, Mg, Si, S, Ar, Ca, Fe and Ni abundances. Ne, Na and Al abundances
are tied to O; metal abundances for a given species of one component
are tied to those of the others. The five temperatures, which are
fixed, are kT$_1=0.5$ keV, kT$_2=1.0$ keV, kT$_3=2.0$ keV,
kT$_4=4.0$ keV and kT$_5=8.0$ keV, the power-law index, as for the
2TP model, is fixed to 1.8.  Throughout our analysis we adopt solar units
according to \citet{anders89}.

Spectral fits are conducted in the 0.5-11 keV band for both \mos and
\pnn. The low energy cutoff is enforced to avoid residual
calibration \citep{kirsch07} and cross-calibration \citep{stuhlinger08}
problems, moreover the relatively high column density towards the
Perseus cluster reduces significantly the value of data accumulated
at the softest energies.


%

\subsection{Analysis in rings} \label{sec: anal_ring}
We investigate a region within 3$^\prime$ from the emission peak,
where evidence of non-thermal emission has been found with \chandra
\citep[and references therein]{sand07}. We consider spectra in 3
annuli with bounding radii: 0.5$^\prime$-1.0$^\prime$,
1.0$^\prime$-2.0$^\prime$ and 2.0$^\prime$-3.0$^\prime$. We exclude
the region within 0.5$^\prime$ from the emission peak to avoid
contamination from the AGN.

From the analysis of the \chandra data \citep[and references
therein]{sand07} we expect the putative non-thermal component to
make up no more than 10-20\% of the total emission, even at high
energies where its relative contribution is largest. Under such
conditions a meticulous analysis of the data is mandatory. We begin
by considering  the annulus with bounding radii
1$^{\prime}$-2$^{\prime}$.

\subsubsection{The 1$^{\prime}$-2$^{\prime}$ ring} \label{sec: ring_1_2}

\begin{figure}
  \centering
  \resizebox{80mm}{!}{\includegraphics[angle=-90]{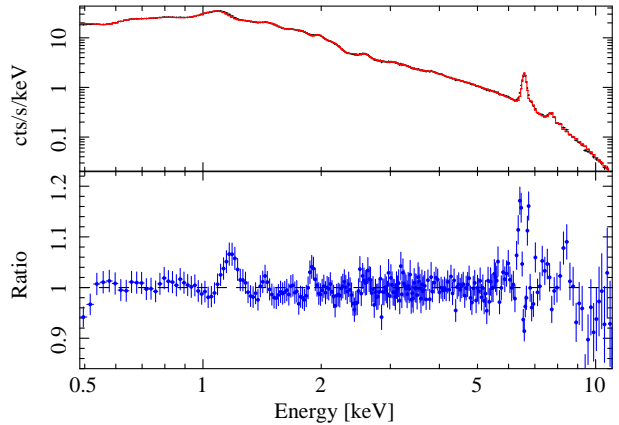}} \\
  \caption{Best fit with the 2TP model to the \pn spectrum in the 1$^{\prime}$-2$^{\prime}$ ring.
  In the top panel we show the spectrum, in black, and the best fitting model in red.
  In the bottom panel we show residuals in the form of a ratio of the data over the model. }
  \label{fig: pnspec}
\end{figure}

A fit to the \pn data with the MTP model returns values of the
surface brightness of the non-thermal component, $ S_{NT} $, of
$(1.9\pm 0.3)\times 10^{-16}$erg$~$cm$^{-2}$s$^{-1}$arcsec$^{-2}$,
the 2TP model returns a similar value of $(1.6\pm 0.3)\times
10^{-16}$erg$~$cm$^{-2}$s$^{-1}$arcsec$^{-2}$  (see also
Tab.~\ref{tab: pnfits}). Interestingly a fit of the MTP and 2TP
models on the \pn spectrum with no systematic errors applied,
hereafter MTP-nosys and  2TP-nosys, returns larger values for the
non-thermal component surface brightness (see Tab.~\ref{tab:
pnfits}).

Inspections of the best-fit residuals reveals three major features,
namely:  an excess around 1.2 keV; a substantial structure around
the Fe K$\alpha$ line and an excess around 8.3 keV (see
Fig.~\ref{fig: pnspec}). The 1.2 keV excess is located in the blue
wing of the Fe L-shell blend and is most likely due to an inaccurate
modeling of the L-shell transitions within the mekal code. Indeed
fitting with the apec rather than the mekal code results in a
substantial reduction of the residuals, with no significant
modification of the best fit parameters.
The excess around 8.3 keV is due to background subtraction
inaccuracies. We remind our readers that in the \pn camera the
region around 8 keV is populated by fluorescence lines of
instrumental origin which are not easily subtracted. Fortunately the
contamination in the 1$^{\prime}$-2$^{\prime}$ region is only mild
\citep{ehle07} and the ensuing  systematics are relatively modest. A
further mitigating factor is the relatively large statistical error
associated to the source spectrum above $\sim 7$keV. The structure
around the Fe K$\alpha$ line is due to an incorrect modeling of the
\pn spectral resolution within the redistribution matrix.
From the analysis of \pn spectra from a sample of clusters \citep[in
preparation]{degrandi08} we have found that the above resolution
miss-calibration can be compensated for by including a
multiplicative component that performs a gaussian smoothing of the
spectral model (\verb|gsmooth| in XSPEC). We set the width of the
gaussian kernel to be 4 eV (FWHM) at 6 keV and assume a power-law
dependency of the width on the energy with an index of $-1$. Both
the MTP and 2TP models modified with the \verb|gsmooth| component,
hereafter MTP-gs and 2TP-gs respectively, provide a substantially
better fit than their non-modified counterparts ($\Delta \chi^2 \sim
200$). The surface brightness of the non-thermal component is
essentially unchanged with respect to the previous fits both for
the MTP-gs, $ S_{NT} = (1.7\pm 0.7)\times 10^{-16}$erg$~$cm$^{-2}$s$^{-1}$arcsec$^{-2}$
and 2TP-gs model $ S_{NT} = (1.5 \pm 0.8)\times 10^{-16}$erg$~$cm$^{-2}$s$^{-1}$arcsec$^{-2}$.

\begin{table}
  \caption{Parameters from the \emph{pn} 1$^\prime$-2$^\prime$ spectrum  fits
  }
  \label{tab: pnfits}
  \centering
  \begin{tabular}{l|l|l|l|l|l}
    \hline \hline
    Model      & $Z_{Fe}^a$     & $Z_{Fe}^a$     & $S_{NT}^{b}$  & $\chi^2$ & dof \\
    \hline
    2TP-nosys  & $0.53\pm 0.01$ & ---            & $4.3\pm 0.3 $ &1540      & 509   \\  
    MTP-nosys  & $0.53\pm 0.01$ & ---            & $3.7\pm 0.5 $ &1489      & 508   \\
    \hline
    2TP        & $0.48\pm 0.01$ & ---            & $1.6\pm 0.3 $ & 636      & 509   \\  
    MTP        & $0.50\pm 0.01$ & ---            & $1.9\pm 0.3 $ & 631      & 508   \\
    \hline
    2TP-gs     & $0.50\pm 0.01$ & ---            & $1.5\pm 0.8 $ & 453      & 509   \\  
    MTP-gs     & $0.52\pm 0.01$ & ---            & $1.7\pm 0.7 $ & 449      & 508   \\  
    \hline
    2TP-gs-2Fe & $0.63\pm 0.06$ & $0.47\pm 0.02$ & $1.1\pm 0.8 $ & 447      & 508   \\  
    MTP-gs-2Fe & $0.79\pm 0.10$ & $0.49\pm 0.01$ & $1.3\pm 0.7 $ & 437      & 507   \\
   \hline
  \end{tabular}
  \begin{list}{}{}
    \item[Notes:] $^a$ Fe abundance in solar units according to \citet{anders89}.
                  $^b$ Surface brightness in units of
                  $10^{-16}$erg$~$cm$^{-2}$s$^{-1}$arcsec$^{-2}$ in
                  the 2-10 keV band.
    \end{list}
\end{table}

A further modification concerns the treatment of the Fe abundance.
When fitting multi-phase models it is customary to tie the
abundances of the different phases; this solution is adopted because
in most cases it is impossible to derive independent metal abundance
measures. In the case at hand it is possible to have two largely
independent measurements of the Fe abundance, one for the hot phase,
contributing mainly to the K-shell line emission, and the other for
the cool phase dominating the L-shell blend emission. We recall
that, for most cool core clusters, spatially resolved analysis shows
that the Fe abundance anti-correlates with temperature (i.e. Fig. 11
of \citealt{sand04} for the Perseus cluster). We have modified our
two test models to allow for different Fe abundances in the cool and
the hot phase. For the 2TP-gs model we simply decouple the Fe
abundance of the two components while for the MTP-gs model we define
two Fe abundances:  one for the cool i.e. 0.5, 1.0 and 2.0 keV
components and the other for the hot, i.e.  4.0 and 8.0 keV
components. As expected for both  models, hereafter 2TP-gs-2Fe and
MTP-gs-2Fe, the Fe abundance for the cool phase or phases is higher
($Z_{Fe} = 0.63 \pm 0.06 $ and $Z_{Fe} = 0.79 \pm 0.10$
respectively) than for the hot phase ($Z_{Fe} = 0.47\pm 0.02 $ and
$Z_{Fe} = 0.49 \pm 0.01$ respectively). Both the MTP-gs-2Fe and
2TP-gs-2Fe models provide a better fit than their single Fe
abundance counterparts although the improvement is not a dramatic
one ($\Delta \chi^2 \sim 10$). The surface brightness of the
power-law component is found to be $ S_{NT} = (1.1\pm 0.8)\times
10^{-16}$erg$~$cm$^{-2}$s$^{-1}$arcsec$^{-2}$ for the 2TP-gs-2Fe
model and $ S_{NT} = (1.3 \pm 0.7) \times
10^{-16}$erg$~$cm$^{-2}$s$^{-1}$arcsec$^{-2}$ for the MTP-gs-2Fe
model. In retrospect it is fairly easy to understand why the
decoupling of hot and warm phase Fe abundances entails somewhat
lower intensities for the non-thermal component. A high temperature
component ($kT \sim$ 8 keV) over a small energy range is rather
similar to a power-law component, the major discriminant being
of-course the line emission. If the high temperature component is
forced to have a metal abundance that is larger than the true one it
will not be able to reproduce the continuum correctly, such an
excess continuum will be well modeled by a power-law component.

We may summarize our results by stating that while the addition of
systematic errors to the spectrum
results in a reduction of the intensity of the power-law component of
a factor of 2, the introduction of some modifications to our
base-line models 2TP and MTP, either to improve the description of
the astrophysical scenario or to account for residual calibration
problems, does not result in substantial modifications of the
intensity of the non-thermal component.

To asses how different choices for the power-law slope impact on the
estimate of the surface brightness of the non-thermal component we
have rerun fits with slopes of $\Gamma=1.5 $ and $\Gamma=2.0$. We
find that surface brightnesses are somewhat smaller (larger) for
$\Gamma=1.5$ ($\Gamma=2.0 $). For instance, in the case of the
MTP-gs-2Fe model, the surface brightness is $ S_{NT} = (1.2\pm
0.8)\times 10^{-16}$erg$~$cm$^{-2}$s$^{-1}$arcsec$^{-2}$ and $
S_{NT} = (2.0\pm 0.8)\times
10^{-16}$erg$~$cm$^{-2}$s$^{-1}$arcsec$^{-2}$ respectively for the
$\Gamma=1.5$ and $\Gamma=2.0 $ choices.
To gauge the effect of systematic errors on our fitting procedure
we have experimented with values ranging from 0\% to 4\%.
For the MTP model we find that the surface brightness of the
non-thermal component steadily decreases from
$( 3.7 \pm 0.5)\times 10^{-16}$erg$~$cm$^{-2}$s$^{-1}$arcsec$^{-2}$
for no systematic error to
$( 1.2 \pm 0.3)\times 10^{-16}$erg$~$cm$^{-2}$s$^{-1}$arcsec$^{-2}$
for a 4\% systematic error.
To verify whether spatial resolution plays a role in the
determination of the power-law surface brightness we have analyzed
spectra from circular sectors. For the 1-2 arcmin annulus we
consider 8 equally spaced sectors each 45 degrees wide.
Analysis of the sectors provides results similar to those found in rings
albeit with a somewhat reduced statistics.


\subsubsection{The \mos spectra for the 1$^{\prime}$-2$^{\prime}$ ring} \label{sec: ring_1_2_mos}

As anticipated we have used \mn and \md spectra to verify \pn
results. A fit with the simple MTP and 2TP models returns values of
the surface brightness of the non-thermal component between 3 and 4
$\times10^{-16}$erg$~$cm$^{-2}$s$^{-1}$arcsec$^{-2}$, (see
Tabs.~\ref{tab: m1fits} and ~\ref{tab: m2fits} for details).
 These values are in agreement with each other and somewhat in excess of
those obtained with the \pnn. Fits on the \mn and \md spectra with
no systematic errors applied, dubbed 2TP-nosys and MTP-nosys,
resulted in a higher value of the $\chi^2$; the surface brightness
of the non-thermal component is also generally higher, however this
not always the case
 (see Tabs.~\ref{tab: m1fits} and ~\ref{tab: m2fits} for details).
Decoupling of the hot and cool phase Fe abundances results in
changes similar to those observed for the \pnn: the $\chi^2$ suffers
a modest decrease; the hot phase Fe abundance is found to be smaller
than the cool phase Fe abundance and the surface brightness of the
non-thermal component does not vary substantially. Inspection of the
residuals shows that the \mn and \md Fe K$\alpha$ lines are not
broader than the one included in the best fitting model,
consequently we refrain from applying a smoothing component to the
spectral model. Conversely we find evidence of some residual
non-linearity in the energy scale calibration, particularly in the
\md spectrum. We have compensated for this effect by allowing for a
constant shift in the energy scale, this is achieved within XSPEC by
performing a \verb|gain fit| on the best fit model, fixing the
linear parameter to 1 and leaving the offset parameter free. From
the resulting best fits, which we refer to as 2TP-2Fe-gf and
MTP-2Fe-gf, we detect a shift of about 2 eV for \mn and 6 eV for
\mdn, other best fit parameters are not substantially different from
those measured from the previous set of fits (see Tabs.~\ref{tab:
m1fits} and ~\ref{tab: m2fits} for details).

\begin{table}
  \caption{Parameters from the \mn 1$^\prime$-2$^\prime$ spectrum  fits
  }
  \label{tab: m1fits}
  \centering
  \begin{tabular}{l|l|l|l|l|l}
    \hline \hline
    Model      & $Z_{Fe}^a$     & $Z_{Fe}^a$     & $S_{NT}^{b}$  & $\chi^2$ & dof \\
    \hline
    2TP-nosys  & $0.66\pm 0.02$ & ---            & $5.4\pm 1.0 $ & 951      & 553   \\
    MTP-nosys  & $0.63\pm 0.01$ & ---            & $4.2\pm 0.5 $ &1009      & 552   \\
     \hline
    2TP        & $0.62\pm 0.02$ & ---            & $3.8\pm 1.1 $ & 593      & 553   \\
    MTP        & $0.59\pm 0.01$ & ---            & $3.0\pm 0.9 $ & 599      & 552  \\
    \hline
    2TP-2Fe    & $0.67\pm 0.04$ & $0.53\pm 0.05$ & $2.7\pm 1.4 $ & 591      & 552   \\
    MTP-2Fe    & $0.66\pm 0.08$ & $0.58\pm 0.02$ & $3.0\pm 0.9 $ & 598      & 551   \\
    \hline
    2TP-2Fe-gf & $0.67\pm 0.05$ & $0.54\pm 0.05$ & $2.7\pm 1.5 $ & 611      & 551   \\
    MTP-2Fe-gf & $0.71\pm 0.12$ & $0.58\pm 0.02$ & $2.4\pm 1.3 $ & 592      & 550   \\
   \hline
  \end{tabular}
  \begin{list}{}{}
    \item[Notes:] $^a$ Fe abundance in solar units according to \citet{anders89}.
                  $^b$ Surface brightness in units of
                  $10^{-16}$erg$~$cm$^{-2}$s$^{-1}$arcsec$^{-2}$ in
                  the 2-10 keV band.
    \end{list}
\end{table}

\begin{table}
  \caption{Parameters from the \md 1$^\prime$-2$^\prime$ spectrum  fits
  }
  \label{tab: m2fits}
  \centering
  \begin{tabular}{l|l|l|l|l|l}
    \hline \hline
    Model      & $Z_{Fe}^a$     & $Z_{Fe}^a$     & $S_{NT}^{b}$  & $\chi^2$ & dof \\
    \hline
    2TP-nosys  & $0.63\pm 0.01$ & ---            & $7.4\pm 0.4 $ &1066      & 555   \\
    MTP-nosys  & $0.56\pm 0.01$ & ---            & $2.6\pm 0.5 $ &1228      & 554   \\
     \hline
    2TP        & $0.61\pm 0.02$ & ---            & $4.2\pm 1.5 $ & 635      & 555   \\
    MTP        & $0.55\pm 0.01$ & ---            & $3.4\pm 0.7 $ & 657      & 554  \\
    \hline
    2TP-2Fe    & $0.65\pm 0.04$ & $0.37\pm 0.10$ & $2.8\pm 1.5 $ & 632      & 554   \\
    MTP-2Fe    & $0.71\pm 0.07$ & $0.52\pm 0.03$ & $2.9\pm 0.8 $ & 655      & 553   \\
    \hline
    2TP-2Fe-gf & $0.74\pm 0.13$ & $0.49\pm 0.07$ & $3.6\pm 1.7 $ & 611      & 553   \\
    MTP-2Fe-gf & $0.80\pm 0.15$ & $0.52\pm 0.02$ & $2.9\pm 1.5 $ & 615      & 552   \\
   \hline
  \end{tabular}
  \begin{list}{}{}
    \item[Notes:] $^a$ Fe abundance in solar units according to \citet{anders89}.
                  $^b$ Surface brightness in units of
                  $10^{-16}$erg$~$cm$^{-2}$s$^{-1}$arcsec$^{-2}$ in
                  the 2-10 keV band.
    \end{list}
\end{table}

As for \pn spectra we have experimented with different values for
the power-law slopes,  the systematic error and performed analysis of spectra in sectors.
Our findings are similar to those reported for the \pn spectra in as far
as the power-law slope and analysis in sectors is concerned.
There are some differences regarding systematic errors. For the \md detector,
unlike for the \mn and \pn detectors, inclusion of a systematic error
results in a reduction of the non-thermal component surface brightness only
for values of the error larger than 3\%.

\subsubsection{Comparison between \pn and \mos estimates} \label{sec: pn_vs_mos}

Comparing results for \mn and \md with those from \pn we find that
the surface brightness of the non-thermal component gauged with the
former experiments is somewhat larger than that measured with the
latter, however, for most measures, the differences are comparable
to the errors associated to individual measurements. More
specifically, if we exclude measurements without  correction for
systematic errors we find  that  \pn measures are roughly contained between
1$\times10^{-16}$erg$~$cm$^{-2}$s$^{-1}$arcsec$^{-2}$ and
2$\times10^{-16}$erg$~$cm$^{-2}$s$^{-1}$arcsec$^{-2}$ and \mos
measures between
2$\times10^{-16}$erg$~$cm$^{-2}$s$^{-1}$arcsec$^{-2}$ and
4$\times10^{-16}$erg$~$cm$^{-2}$s$^{-1}$arcsec$^{-2}$.

If we consider measurements with no correction
for systematics (i.e. MTP-nosys bst fits) we find that typical
errors, 0.5$\times10^{-16}$erg$~$cm$^{-2}$s$^{-1}$arcsec$^{-2}$,
are substantially smaller than the spread of values described above, indicating
that systematic uncertainties afford the dominant contribution to the error budget.
Under these circumstances providing a statistically meaningful confidence range
for our measurement of the surface brightness of the non-thermal component
is rather difficult, particularly since the approach we have adopted to
incorporate systematic errors is of a heuristic nature.
We therefore resort to an alternative and less ambitious approach: we make use of the
spread of values measured for the different models and different experiments to
provide a loose range for the surface brightness of the non-thermal component, namely 0-5$\times10^{-16}$erg$~$cm$^{-2}$s$^{-1}$arcsec$^{-2}$.
It is perhaps ironic that a meticulous analysis of the data, such as the one reported here,
has lead to an uncommonly loose estimate of the non-thermal component, however,
as long as we do not dispose of statistical tools allowing us to incorporate systematic
uncertainties in analysis such as these, our options are rather limited.



The results we report here are at variance with those that we
presented at a conference in Garching in August 2006, essentially 4
factors have lead to these differences. The factors are: a new
release of the \mos quantum efficiency curves in August 2007 leading
to a substantial improvement of the \mos vs. \pn cross-calibration
and to a revision of \mos spectral analysis results; use of
\verb|PATTERN|==0  events only for \mos spectra; a more accurate
treatment of the background following some recent work within our
group \citep{leccardi07,leccardi08}; a novel treatment for
systematic effects.

\subsection{Radial profile} \label{sec: rad_prof}

\begin{figure}
  \centering
  \resizebox{75mm}{!}{\includegraphics[angle=-90]{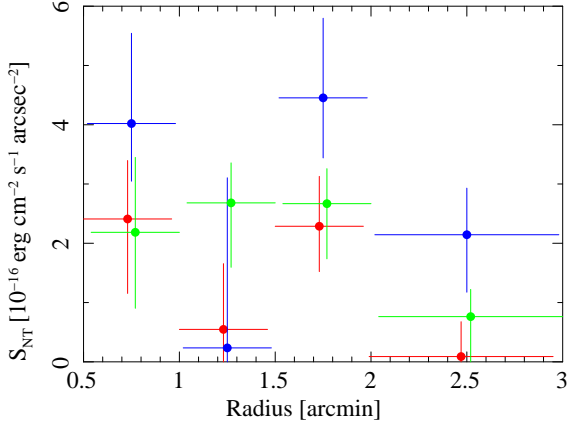}} \\
  \caption{Radial profile for the surface brightness of the non-thermal component for the three
  \epic cameras: \pn is red, \mn is blue and \md is green. Radii have been slightly offset to
improve readability.}
  \label{fig: rad_prof}
\end{figure}

In Fig.~\ref{fig: rad_prof} we show the radial profile for the
surface brightness of the non-thermal component. The profile is
based on the analysis of spectra accumulated in annuli with bounding
radii: 0.5$^\prime$-1.0$^\prime$, 1.0$^\prime$-1.5$^\prime$,
1.5$^\prime$-2.0$^\prime$ and 2.0$^\prime$-3.0$^\prime$, note that
the 1.0$^\prime$-2.0$^\prime$ annulus has been split in two to
improve the resolution of the profile. The reported surface
brightnesses have been estimated using the MTP-gs-2Fe model for \pn
and MTP-2Fe-gf model for \mn and \mdn.

All measurements lie between 0 and
5$\times10^{-16}$erg$~$cm$^{-2}$s$^{-1}$arcsec$^{-2}$.
For the innermost ring
the \epic mean value for the surface brightness of the non-thermal
component is $(2.4\pm 1.0
)\times10^{-16}$erg$~$cm$^{-2}$s$^{-1}$arcsec$^{-2}$, while for the
1.5$^\prime$-2.0$^\prime$ annulus we find $S_{NT} = (2.3\pm 0.9 )
\times10^{-16}$erg$~$cm$^{-2}$s$^{-1}$arcsec$^{-2}$.

 We have verified that similar measurements are obtained when fitting
spectra with slightly different spectral models, i.e. all \pn measurements
obtained with the MTP-gs  model are within 1$\sigma$ of those
reported in Fig.~\ref{fig: rad_prof}, similarly all \mn and \md
measures obtained with the MTP-2Fe  model are within 1$\sigma$ of those
reported in Fig.~\ref{fig: rad_prof}.

Assuming a rather conservative approach, we might say that  the surface
brightness of the non-thermal component is somewhere between 0 and
5$\times10^{-16}$erg$~$cm$^{-2}$s$^{-1}$arcsec$^{-2}$. This is
 in agreement with what we found from the analysis of the
1.0$^\prime$-2.0$^\prime$ annulus (see Sect.~\ref{sec: pn_vs_mos})
and may therefore be considered as a general result of our analysis.

\section{Comparison with \chandra} \label{sec: chandra}

In an attempt to understand the origin of the difference between our
measurements and those based on \chandra data, we have reduced and
analyzed the longest \emph{ACIS-S} observation
\citep[see][Tab.~1]{fabian06}. The data were reduced with the X-ray
analysis packages \ciao\ 3.4 and \heasoft\ 6.4 in conjunction with
the \chandra calibration database (\emph{Caldb}) version 3.4.2
(using gainfile acisD2000-01-29gain\_ctiN0006.fits). To ensure the
most up-to-date calibration, all data were reprocessed from the
``level 1'' events files, following the standard \chandra
data-reduction
threads\footnote{{http://cxc.harvard.edu/ciao/threads/index.html}}.
We applied the standard corrections to account for a time-dependent
drift in the detector gain and charge transfer inefficiency, as
implemented in the \ciao\ tools.  From low surface brightness
regions of the active chips we extracted a light-curve
(5.0-10.0~keV) to identify and excise periods of enhanced
background. The spectra were re-binned to ensure a S/N of at least 3
and a minimum 20 counts per spectral bin. Spectral fits, using both
2TP and MTP models provide values consistent with those derived by
\citet{sand07} (see their Fig.~20) and well in excess of
those measured with \epic (see Tab.~4). We have included a 2\%
systematic error to the \chandra spectrum, as done for the \epic
spectra, without noticeable effects on the non-thermal component, or
any other spectral parameter. The background file represents well
the background in the actual observation, as gauged by the 9-13 keV
band count-rate and even if we experiment changing the normalization
by $\pm$5\% the flux in the power-law component is comparable with
the one measured with the background at the nominal value
($7.6\pm0.7$ and $8.9^{+0.5}_{-0.8}$, in the units of Tab.~4 and
using the MTP model, in the two cases). We investigated  possible
variations in the set of Perseus observations reducing other long
observations, namely OBS-ID 4948 and OBS-ID 4289, in particular the
latter provides an important double check because it was performed
at an earlier epoch and with a different level of low-energy
degradation. We did not find any difference within the errors for
the value of the spectral parameters and in particular of the
non-thermal component between the different observations.

\begin{table}
  \caption{Parameters from the \chandra \st 1$^\prime$-2$^\prime$ spectrum  fits
}
  \label{tab: cxofits}
  \centering
  \begin{tabular}{l|l|l|l|l|l}
    \hline \hline
    Model      & $Z_{Fe}^a$     & $Z_{Fe}^a$     & $S_{NT}^{b}$  & $\chi^2$ & dof \\
    \hline
    2TP-nosys  & $0.81\pm 0.01$ & ---            & $10.3^{+0.4}_{-0.3} $ & 2358      & 522   \\
    MTP-nosys  & $0.79\pm 0.01$ & ---            & $8.8^{+0.3}_{-0.2}$ & 2360      & 521   \\
    \hline
    2TP        & $0.73^{+0.03}_{-0.02}$ & ---            & $8.4^{+0.8}_{-0.6} $ & 589      & 522   \\
    MTP        & $0.74^{+0.03}_{-0.02}$ & ---            & $8.6^{+0.5}_{-1.0} $ & 584      & 521   \\
    \hline
    2TP-2Fe    & $0.95^{+0.10}_{_0.08}$ & $0.68\pm 0.03$ & $7.8^{+1.3}_{-0.7} $ & 582      & 521   \\
    MTP-2Fe    & $1.20^{+0.65}_{-0.13}$ & $0.67^{+0.03}_{-0.02}$ & $7.1^{+1.3}_{-0.7} $ & 578      & 520   \\
   \hline
  \end{tabular}
  \begin{list}{}
    \item[Notes:] $^a$ Fe abundance in solar units according to \citet{anders89}.
                  $^b$ Surface brightness in units of
                  $10^{-16}$erg$~$cm$^{-2}$s$^{-1}$arcsec$^{-2}$ in
                  the 2-10 keV band.
    \end{list}
\end{table}

\begin{figure}
  \centering
  \resizebox{75mm}{!}{\includegraphics[angle=-90]{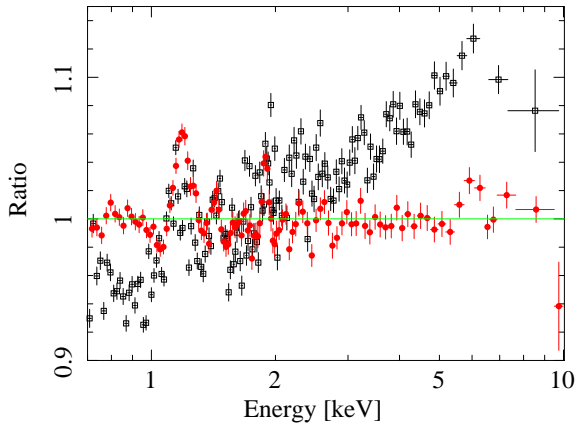}} \\
  \caption{Ratio of the \st (black) and \pn (red) spectra accumulated in the
   1$^{\prime}$-2$^{\prime}$ annulus over the best fitting MTP-gs-2Fe \pn model.
  A re-normalization constant of 4.8\% has been applied to the \st spectrum.}
  \label{fig: cxo_pn}
\end{figure}

In an attempt to understand the difference between \chandra and
\xmmn measurements, we have compared the \epicn-\pn spectrum
accumulated in the 1$^{\prime}$-2$^{\prime}$ annulus with the
\chandra spectrum extracted from the same region. Technically this
has been done by plotting the \pn and the \st  residuals with
respect to the \pn MTP-gs-2Fe best-fit model in the form of a ratio
of the data over the model. A 5\% re-normalization has been applied
to the \chandra spectrum to facilitate the comparison with the \pn spectrum.
Interestingly the \st spectrum  is significantly
harder than the \epicn-\pn spectrum. As shown in Fig.~\ref{fig:
cxo_pn}, while at 2 keV the two spectra have by construction the
same intensity, at 5 keV the \chandra spectrum is in excess of the
\epicn-\pn spectrum by about 10\%. The discrepancy is rather
significant, considering that the non-thermal component intensity is
of the same order.
A comparison of the \pn spectrum with the  \mn and \md
spectra similar to the one reported in Fig.~\ref{fig: cxo_pn} shows that
in the 1-7 keV range both \mn and \md spectra are within a few \% of the \pn
spectrum. We refrain from  plotting the \mos spectra in Fig.~\ref{fig: cxo_pn}
simply to avoid overcrowding.

Since the background of the \st and \epic cameras at energies below
5 keV is extremely small (see Fig.~\ref{fig: cxo_epc}), the most
likely origin for the difference is a cross-calibration issue on the
effective areas in the 1-5 keV range. Interestingly
cross-calibration activities conducted by a number of workers
\citep[see][]{david07}, mostly on cluster spectra, also point to a
discrepancy between \chandra and \xmmn measurements. For hot
clusters, (kT $\approxgt 5$keV) \chandra spectra typically return
higher temperatures than \epic spectra \citep[see][figure on page
2]{david07}. Further investigations have revealed that the
discrepancy likely results from problems in the calibration of the
\chandra high energy effective areas.  \citet{david07}  show that
measuring the temperature of A2029 from the ratio of the He-like
over H-like K$\alpha$ lines and comparing the resulting continuum
model with the observed spectrum, the latter is flatter than the
former. More specifically, if the model is normalized to the
observed spectrum at energies larger than 5 keV,  the latter is
found to be about 8\% below the model at 2 keV. This discrepancy is
remarkably similar to the one we find when comparing the \pn and \st
spectrum of Perseus (see Fig.~\ref{fig: cxo_pn}).

\section{Discussion} \label{sec: discussion}


From a general perspective  our analysis may be viewed as
 an attempt to characterize the spectrum of an X-ray source
 beyond a simple one/two component model. Given the modest
intensity of the  component we are specifically interested in and
the lack of spectral signatures, such as lines or edges, the task is
by no means a trivial one. Clearly it can only be attempted on
sources where adequate statistics is available. It is often the case
that, when the statistics is abundant, systematics become the
dominant source of error. At the present time X-ray astronomers do
not have a standard method of analyzing data under these
predicaments and mostly resort, has we have done, to  a trial and
error approach. The development of a standard strategy, possibly
descending from first principles, is highly desirable particularly
if we consider that in the not too distant future X-ray missions
such as \emph{XEUS} or \emph{Con-X} will have, as one of their primary goals, the
detailed characterization of relatively bright X-ray spectra.
In some instances the advent of high
resolution spectrometry will provide a valid solution. In others,
where features that need to be characterized are broad, either
because they are of  non-thermal nature, such as the putative hard
tails in clusters, or because they are smeared by other processes
(i.e. the broad iron line observed in nearby AGN), solutions will
have to be found elsewhere. Recent work by \citet{drake07} provides
an interesting starting point.

\subsection{\epic estimate} \label{sec: dis_epic}

In this paper we have characterized  the surface brightness of the
non-thermal component in Perseus  by  resorting to a heuristic
approach which can be divided into 2 steps:
 1) inclusion of 2\% systematic errors on spectra;
 2) modification of spectral models.
As far as the first step is concerned, we find a significant
reduction of the surface brightness measured from the \pn spectra
when the 2\% systematic error is included, no significant changes
are found for the \mos measurements ( note however that a reduction is observed
starting from a 3\% systematic error, see Sect.~\ref{sec: ring_1_2_mos} ).
As far as the second step is
concerned, the stability of our results with respect to the
inclusion of corrections for some imperfections in redistribution
matrix, effective area, energy scale and  choice of  astrophysical
model indicate that our measures are relatively robust with respect
to the residual systematic effects we have been able to identify.
Comparison of \mos with \pn results shows that the former favour a
somewhat larger value for the surface brightness of the non-thermal
component than the latter, roughly speaking 2 against 1.  The
difference between the two measures is most likely  caused by minor
cross-calibration issues, possibly in the high energy
response of either \mos or \pnn. Since we do not have firm evidence as to
which of the two experiments is better calibrated, rather than privileging measures
from one over the other, we  take the common envelope of \mos and \pn
measurements (i.e. 0- 5$\times10^{-16}$erg$~$cm$^{-2}$s$^{-1}$arcsec$^{-2}$)
as our best estimate for the interval constraining the surface brightness of
the non-thermal component. Some of
our readers might find this range rather broad, particularly in
light of the high statistical quality of our data, we reiterate that
the major source of indetermination here as elsewhere are systematic
and not statistical errors.

\subsection{\epic vs. \acis } \label{sec: dis_acis}

Results from the analysis of \epic spectra, either \pn or \mosn,
cannot be reconciled with those obtained with \stn. As discussed in
Sect.~\ref{sec: chandra} the difference is to be ascribed to a
cross-calibration issue between \epic and \acisn. Similar problems
have been identified by a group of calibrators who have compared
\chandra and \xmmn observations of hot clusters \citep{david07}.
 Recent efforts by \chandra calibrators \citep{david07,marshall08} have shown that the above
difference follows  from problems in the calibration of the \chandra
high energy effective areas.

\subsection{Magnetic field estimates} \label{sec: dis_b}

 Under the assumption that the non-thermal
emission originates from inverse Compton (IC) scattering of seed
microwave and infrared photons by relativistic electrons,
responsible of the synchrotron emission in the mini radio-halo, the
Perseus core magnetic field can be estimated from the energy density
of the seed photon field, $U_{rad}$, the synchrotron and
inverse-Compton luminosities, $L_R$ and $L_X$, i.e. $ U_B = U_{rad}
\cdot L_R / L_X $, where $ U_B$ is the energy density of the
magnetic field. \citet{sand05}, from their measures of the
non-thermal component, find magnetic field intensities ranging from
several $\mu$G at the very center, to $\sim$ 1.0~$\mu$G at 8 kpc
(0.4$^\prime$) and $\sim$ 0.1~$\mu$G at 40 kpc (2$^\prime$)
\citep[see Fig.~9 of][]{sand05}. Our measurements can be used to
provide a revised estimate of the B field.
As discussed above we estimate the surface brightness of the non-thermal
 component to be somewhere between
0 and 5$\times 10^{-16}$erg$~$cm$^{-2}$s$^{-1}$arcsec$^{-2}$.  We
therefore start by assuming an upper limit of 5$\times 10^{-16}$erg$~$cm$^{-2}$s$^{-1}$arcsec$^{-2}$,
 from this we subtract a value of 2.4$\times 10^{-16}$erg$~$cm$^{-2}$s$^{-1}$arcsec$^{-2}$,
as in \citet{sand05}, to account for projection  effects. The resulting surface brightness is used to
derive lower limits for the magnetic field using essentially the same method and
radio measures described in \citet{sand05}. In the 0.5$^{\prime}$-1$^{\prime}$ and
1$^{\prime}$-2$^{\prime}$ annuli our X-ray estimate
convert into lower limits of roughly 0.4~$\mu$G and 0.3~$\mu$G
respectively.
 We note that these numbers have been derived
through a series of rather drastic approximations, (i.e. we have a rather
poor determination of the X-ray upper limit, the radio emission is measured at a frequency which
is substantially larger than that at which non-thermal electrons responsible
for the inverse Compton emission emit via synchrotron, the B field is
assumed to be constant, etc.)
what really matter is that they are lower limits and not detections.

Faraday rotation measures (RM) in cool cores on scales of tens of
kpc are in the order of several hundreds to thousands
\citep{taylor02}. Estimates of magnetic fields from rotation
measures have undergone some revision in the last few years with
more recent estimates typically in the order of a few $\mu$G
\citep{clarke04,ensslin06}. In the case of the Perseus mini
radio-halo  Faraday rotation measures are available only on very
small scales \citep{taylor06}, i.e. few tens of pc. RM estimates are
in the order of $\sim$7000 rad m$^{-2}$ leading to B field values of
$\sim 25~\mu$G under the assumption the screen is localized in the
ICM. This, however, appears to be unlikely as variations  of 10\% in
the RM are observed on scales of $\sim$ 1 pc \citep{taylor06}, while
ICM magnetic fields are expected to be ordered on significantly
larger scales
 \citep[few kpc:][]{taylor02,vogt05,ensslin06}.  Application of the
 classical minimum-energy argument to the Perseus mini radio-halo data,
leads to estimates for the central (i.e. $r=0$) magnetic field strength of  
$\sim 7~\mu$G \citep{pfrommer04}.

IC estimates of the magnetic field based on \chandra
measurements \citep{sand05}  are about an order of magnitude below
the RM and minimum-energy estimates detailed above. A similar
discrepancy has been found when comparing magnetic field estimates
on cluster wide scales. There IC measures are in the 0.2-1~$\mu$G
 range \citep[and references therein]{rephaeli08}, while RM estimates are
about an order of magnitude larger
\citep[e.g.][]{carilli02,rephaeli08}. While the intricate nature of
the astrophysical scenario lends itself to various possible
explanations for the discrepancy \citep[e.g.][]{carilli02,govoni04},
a more trivial alternative, namely that the IC estimates might be
incorrect, should also be considered. There is at least one object,
Coma, where IC magnetic field measures
\citep{fusco99,fusco04,fusco07} have been challenged
\citep{rossetti04,rossetti07}. Moreover first measures from the
\emph{Suzaku} mission on a few objects, (i.e. A3667, Coma) are
turning out to be  lower limits on IC measures \citep[and references
therein]{fuka07}. \xmmn measures presented in this paper seem to
play a similar role in the determination of  the Perseus core IC
magnetic field estimates.




\subsection{Future prospects} \label{sec: dis_future}

It is unlikely that either \chandra or \xmmn  will provide
significantly better estimates of the intensity of the non-thermal
component in Perseus than those reported here. Clearly experiments
with sensitivities extending into the hard X-ray band are the most
appropriate for these kind of studies.
 Recently the \emph{BAT} experiment on board the \emph{Swift} satellite
has been used to determine that an extrapolation of the non-thermal flux
measured with  \chandra to the 50-100 keV band overshoots the flux
measured with \emph{BAT} by a factor of about 4 \citep{ajello08}. 
The \emph{Suzaku} high energy
experiment \citep{taka07}, limited as it is by modest spatial resolution and
by a non optimal background treatment \citep{kokubun07}, may provide some useful
indications but will most likely not allow substantial advancement.
Amongst missions that are currently under development \emph{Simbol-X}
\citep{ferrando05}
is arguably one of the most promising. The combination of large throughput
in the 1-60 keV range, low instrumental background and \epicn-like
spatial resolution, will allow a sensitive measurement of non-thermal
components extending beyond the thermal cutoff. We have performed
simulations of \emph{Simbol-X} spectra based on  our best fits for the
1$^{\prime}$-2$^{\prime}$ annulus trying out different values of the
normalization of the non-thermal component to determine how far down
we might go. We started by taking a value of the surface brightness
of 1$\times10^{-16}$erg$~$cm$^{-2}$s$^{-1}$arcsec$^{-2}$. We find
that the non-thermal component dominates thermal emission above
$\simeq$ 20 keV and remains above background emission up to $\simeq$
30 keV. Thus, if the non-thermal emission has an intensity of this
magnitude or larger it will be detected rather easily. Repeating the
same exercise with a surface brightness that is ten times smaller,
i.e. 1$\times10^{-17}$erg$~$cm$^{-2}$s$^{-1}$arcsec$^{-2}$, we find
that formal fitting of the simulated spectrum still allows a
detection of the non-thermal component at the $\sim 3-4 \sigma$
level, provided the observation is longer than $3\times 10^5$s and
current estimates of the \emph{Simbol-X} instrumental background are
within  10-20\% of the real value. We
note that in this case the relative intensity of the non-thermal
component in the 20-30 keV band is about 10\%.  As we have learned
from the analysis of our \epic data, at these intensity levels,
systematics, which are at this time unknown for \emph{Simbol-X}, will play
an important, possibly dominant role. In the specific case of
\emph{Simbol-X} measurements, the critical elements which need to be kept
under control at the few percent level are: the effective area
calibration; the cross-calibration between the high and low energy
detectors and the background. These requirements, albeit
challenging, are not beyond reach, particularly if we consider that
the formation flight strategy adopted by \emph{Simbol-X} will entail some
important advantages over previous missions. Extensive calibration
of the telescope effective areas can be performed in flight by
comparing observations of calibration sources performed with and
without the telescope in the optical path. Moreover direct
illumination of the \emph{Simbol-X} focal
plane, with the telescope removed, will allow an in-flight
verification of the detector quantum efficiency.

Assuming that \emph{Simbol-X} can indeed reach a sensitivity of
1$\times10^{-17}$erg$~$cm$^{-2}$s$^{-1}$arcsec$^{-2}$ this will
allow to detect IC components with associated B fields of roughly
1~$\mu$G, thereby providing important constraints on the
magnetic field in Perseus.
Given the current magnetic field estimates from Faraday
rotation measures on a few  cool cores \citep[few
$\mu$G:][]{ensslin06} and the  classical minimum-energy argument estimate on
Perseus \citep[$\sim~7\mu$G:][]{pfrommer04}, even the exquisite
sensitivity within reach for \emph{Simbol-X} might be insufficient
to detect the IC emission.


\section{Summary} \label{sec: summary}

We have carried out a detailed analysis of a long \epic observation
of the Perseus core in an attempt to detect and characterize a
non-thermal component, our main findings may be summarized as
follows:

\begin{itemize}
   \item  systematic uncertainties play an important role in the characterization of the non-thermal component;
          in the absence of a strategy descending from first principles we have developed a heuristic
          approach to include them in our analysis;
   \item  at variance with our preliminary estimates, we find that the non-thermal component
          is not detected; the surface brightness is determined to be smaller than $\sim$ 5$\times10^{-16}$erg$~$cm$^{-2}$s$^{-1}$arcsec$^{-2}$;
   \item  our \xmmn estimates are at variance with \chandra estimates from \citet{sand05,sand07} and
          from our own analysis; the most likely explanation for the discrepancy between \chandra and
          \xmmn is a problem in the \chandra effective area calibration;
   \item  our \epic based upper-limit on the surface brightness converts into IC magnetic field lower
          limits of $\sim$0.4~$\mu$G for the 0.5$^{\prime}$-1$^{\prime}$ annulus and $\sim$0.3~$\mu$G for the 1$^{\prime}$-2$^{\prime}$ annulus;
          these measures are  not in disagreement with RM estimates on a few
          cool cores \citep[few $\mu$G:][]{ensslin06} and the minimum energy estimate on Perseus
          \citep[10~$\mu$G:][]{pfrommer04};
   \item in the not too distant future \emph{Simbol-X} may allow detection of non-thermal components with
         intensities more than 10 times smaller than those that can be measured with \epicn; nonetheless
         even the exquisite sensitivity within reach for \emph{Simbol-X} might be insufficient to detect
         the IC emission from Perseus.
\end{itemize}

\begin{acknowledgements}
We would like to express our gratitude to calibrators on both sides
of the Atlantic, without their continuing efforts this work would
not have been possible. We  thank P.J. Humphrey for the use of his
\chandra reduction code, G. Brunetti for useful discussions and M.
Rossetti, S. De Grandi and L. Zappacosta for a critical reading of the manuscript.

\end{acknowledgements}

\bibliographystyle{aa}
\bibliography{refs}

\end{document}